\documentclass[11pt,a4paper]{article}

\usepackage[ansinew]{inputenc}
\usepackage[english]{babel}
\usepackage{amsmath,amsfonts,amssymb,mathrsfs,calc,units,hyperref}
\usepackage[dvips]{graphicx}

\hypersetup{%
linktocpage,colorlinks,urlcolor=blue,citecolor=blue,%
linkcolor=blue,filecolor=blue,%
}

\makeatletter
        \newcommand{\margenes}[4]{%
                \setlength{\oddsidemargin}{\dimexpr #1-\hoffset-1in}                    
                \setlength{\textwidth}{\dimexpr \paperwidth-#1-#2}                      
                \setlength{\topmargin}{\dimexpr #3-\voffset-1in}
                \setlength{\textheight}{\dimexpr \paperheight-#4-#3-\footskip-\headheight-\headsep}
                \if@twoside      \setlength{\evensidemargin}{\dimexpr #2-\hoffset-1in}  \fi }
\makeatother

\margenes{2cm}{2cm}{2cm}{2cm}

\long\def\symbolfootnote[#1]#2{\begingroup%
\def\thefootnote{\fnsymbol{footnote}}\footnote[#1]{#2}\endgroup} 

\begin{document}


\begin{flushright}
\rightline{ICCUB-13-001}

\rightline{February, 2013}
\end{flushright}

\bigskip

\begin{center}

{\Large\bf  
Supersymmetric  BCS:\\[2mm]
{\large Effects of an external magnetic field and spatial fluctuations of the  gap}
}

\bigskip
\bigskip

{\it \large Alejandro Barranco}
\bigskip

{\it\centering
Institute of Cosmos Sciences and ECM Department, Facultat de F{\'\i}sica\\
Universitat de Barcelona, Av. Diagonal 647,  08028 Barcelona, Spain\\[2mm]
Email address: \sf alejandro@ecm.ub.es
}

\end{center}
\bigskip
\bigskip

\begin{abstract}
Recently an $\mathcal N=1$ supersymmetric model of BCS superconductivity was
proposed realizing spontaneous symmetry breaking of a $U(1)_R$ symmetry. Due
to scalar contributions the superconducting phase transition turned out to 
be first order rather than second order as in standard BCS theory. Here we
consider the effects of an external magnetic field and spatial fluctuations 
of the gap in that model. This allows us to compute the magnetic penetration
length and the coherence length, and also to distinguish between type I and
type II superconductors. We compare the supersymmetric and standard
relativistic BCS results, where the main differences come from the different
orders of the phase transition.
\end{abstract}

\clearpage

\tableofcontents


\section{Introduction}

Certain condensed matter systems have scalar and fermionic excitations
showing a quasi-supersymmetric dynamics, for example BCS theory, where Nambu
\cite{Nambu} found a certain quantum mechanical quasi-supersymmetry, in terms of which 
one can describe the Interating Boson Model \cite{Iachello}. 
In the past few years, some attempts to study the supersymmetric generalization
of relativistic BCS theory 
have appeared in the literature \cite{Harnik-Larson-Murayama,Maru-Tachibana,Ohsaku08}. 
However, these models
include explicit supersymmetry breaking terms to stabilize the scalar
sector. In this paper we proceed with the work done in \cite{Barranco-Russo}, 
where a model of supersymmetric BCS superconductivity
was proposed without the need of introducing explicit supersymmetry breaking 
mass terms to avoid Bose-Einstein condensation (BEC) of the scalars.

Apart from being an interesting problem illustrating
the difficulties that the introduction of a chemical potential in a supersymmetric 
theory brings up, implementing the BCS mechanism within the context of a supersymmetric 
field theory might be interesting because some exact non-perturbative results can be 
computed for supersymmetric theories (e.g. \cite{Intriligator-Seiberg}). Therefore, 
this may help to the understanding of strong coupling physics in superconductivity theories.
Moreover, supersymmetric field theories are more stable and less sensitive to radiative corrections,
 as a result, the theory is less sensitive to the UV cutoff,
which in BCS theory must be put by hand by introducing a Debye energy
as a phenomenological input of the model.
Another motivation is the fact that having scalar superpartners for fermions in the supersymmetric 
version of BCS theory might be of relevance for real condensed matter systems where scalar 
and fermionic excitations arise in a quasi-supersymmetric way, like in BCS theory itself, as noticed
by Nambu.

The implementation of BCS superconductivity requires the presence of a Fermi
surface and the introduction of a chemical potential for fermions, but these 
two conditions may lead to problems when scalars are included in the
supersymmetric version. If the chemical potential is introduced through a
baryonic $U(1)_B$ symmetry, supersymmetry demands that both fermions and
scalars have the same coupling to the chemical potential. If the chemical
potential is greater than the mass, a feature which turns out to be necessary
for the existence of a Fermi surface,  scalars undergo Bose-Einstein
condensation which prevents the BCS mechanism to be at work.
Fortunately, there is a $U(1)_R$ symmetry which allows us to introduce the
chemical potential for fermions while keeping the scalars neutral, avoiding in
this way  BEC. The main difference between this supersymmetric model and
standard BCS, is that, as we will explain, IR physics of scalars makes 
the superconducting phase transition first order.
 
Our aim here is to study the supersymmetric model in \cite{Barranco-Russo}
under  the effect of an external magnetic field, which will allow us to
compute the magnetic penetration length and describe the Meissner effect. 
We also consider the effect of spatial fluctuations of the field $\Delta$
around its vacuum expectation value, necessary to compute the coherence
length. After computing these two lengths, we can compute the critical magnetic
fields, $H_{c1}$ and $H_{c2}$, that we would  have for a type-II
superconductor, so that compared with the critical magnetic field of a type I
superconductor, $H_c$, allows us to distinguish between the two types.
All this is done putting emphasis in the comparison between the supersymmetric 
and non-supersymmetric results.

The paper is organized in the following way. In section 2, we review the relativistic
BCS theory and comment on how to obtain the aforementioned characteristic
quantities for a superconductor as functions of the temperature. In section 3, 
we briefly explain  the supersymmetric BCS model. In section 4, we go on with
the comparison of the quantities computed within the two theories, and comment
on whether the superconductors are type I or II. Finally, we end up in section
5 with some conclusions. Details of the computations are given in the
appendices. We will follow the notation of \cite{Terning} in writing the
Lagrangians.

\section{Relativistic BCS theory}\label{sec:2}

In this section we will review the relativistic
BCS theory following the line of reasoning in \cite{Bertrand,Barranco-Russo}.
As shown in \cite{Bertrand}, the relativistic BCS theory can be described 
in terms of a four fermion interaction by the Lagrangian 
\begin{equation}\label{LBertrand}
\mathcal L=\frac{i}{2} (\bar   \psi \gamma^\mu \partial_\mu \psi 
-\partial_\mu \bar  \psi \gamma^\mu \psi)
-m\bar \psi \psi+\mu\psi^\dagger\psi
+ {g^2}(\bar  \psi_c \gamma_5 \psi )^\dagger 
(\bar  \psi_c \gamma_5 \psi )\ .
\end{equation}
This Lagrangian enjoys a global $U(1)$ symmetry\footnote{Strictly speaking, if the symmetry is global,
we are dealing with superfluidity, but transport properties are similar to those of superconductivity
and we can consider this model as that of a superconductor in the limit in which the $U(1)$ is ``weakly
gauged''.}, this allows us to introduce the
chemical potential associated with the corresponding conserved charge. 
BCS is a theory describing a spontaneous symmetry breaking of this $U(1)$ symmetry 
driven by the temperature. For this reason, to study the theory at finite temperature
we have to consider the Lagrangian \eqref{LBertrand} in Euclidean space.
At this point, we have a Lagrangian which is not quadratic in the fields,
to alleviate this, one has to perform a Hubbard-Stratonovich transformation by
introducing an auxiliary field, $\Delta$. At the end of the day we arrive at the Lagrangian
\begin{align}\label{bcsl}
\mathcal L_E &=\frac{1}{2} (\psi^\dagger  \partial_\tau \psi -\partial_\tau \psi^\dagger \psi)
-\frac{i}{2} (\bar   \psi \gamma^i \partial_i \psi -\partial_i \bar   \psi \gamma^i\psi )
+m\bar   \psi \psi-\mu\psi^\dagger\psi
\nonumber\\&\phantom{=\ }
+{g^2} |\Delta|^2 -{g^2} \big[ \Delta^\dagger  (
\bar \psi_c \gamma_5 \psi )+ \Delta (\bar\psi_c \gamma_5 \psi )^\dagger \big]\ .
\end{align} 
In this way, if we eliminate the auxiliary field  through its equations of
motion, we recover the original Lagrangian \eqref{LBertrand}. The equations of motion
set  $\Delta=(\bar\psi_c \gamma_5 \psi)$, where we see that $\Delta$ is a measure of the density
number of Cooper pairs.

Once we have a Lagrangian which is quadratic in fermions, our purpose is to integrate them out
to obtain a one-loop effective potential for the auxiliary field $\Delta$. After performing
the Matsubara thermal sums, the effective potential is given by
\begin{align}\label{Veff}
V_{\rm eff}&={g^2} \vert\Delta\vert^2
\nonumber\\&\phantom{=\ }
-\int \frac{d^3p}{(2\pi)^3} \left(\omega_-(p)+ \omega_+(p)\right)
\nonumber\\&\phantom{=\ }
-\frac{2}{\beta} \int \frac{d^ 3p}{(2\pi)^3} \left(\log(1+e^{-\beta \omega_-(p)})+ \log(1+e^{-\beta \omega_+(p)})\right) \ ,
\end{align} 
where we have written in separate lines, the classical potential, the Coleman-Weinberg potential
and the thermal potential, from top to bottom respectively. The $\omega_\pm$ appearing in the 
effective potential are the energy eigenvalues coming from the Lagrangian \eqref{bcsl}, 
and they are given by
\begin{equation}
\omega_\pm=\sqrt{(\omega_0(\vec p) \pm \mu)^ 2+4g^4\vert\Delta\vert^2}\ ,
\end{equation} 
where $\omega_0\equiv\sqrt{p^2+m^2}$. From the expression of the energy eigenvalues it is
manifest why $\Delta$ is called the gap.

The whole one-loop effective potential is identified with the free energy density in the grand
canonical ensemble. It is UV divergent due to the Coleman-Weinberg contribution, for which we 
must introduce a cut-off, $\Lambda$, appearing here like a ``Debye energy''. Considering
the thermal potential at low temperatures, one can see that the dominant contribution comes from 
the minimum of the energy eigenvalues. Supposing $\mu>0$, this minimum is reached for $\omega_-$ 
(which is identified with the particle contribution) around $\omega_0=\mu$, and contributions from $\omega_+$,
i.e. the anti-particle, can be neglected. The momentum space location of this minimum defines 
a Fermi surface, $p_F^2=\mu^2-m^2$, from which it is clear that the condition $\mu> m$ is required for the
Fermi surface to exist.

The minimum of the effective potential as a function of the gap for each temperature defines
a curve $\Delta(T)$, describing a second order phase transition at a certain critical temperature
$T_c$, below which the fermion condensate appears. This curve can be found by solving the gap equation,
$\partial_\varepsilon V_{eff}=0$, where we have introduced the notation $\varepsilon=\vert\Delta\vert^2$.
Explicitly, the gap equation is
\begin{equation}\label{gapBCS}
1 =\frac{g^2}{\pi^2}\int_0^\Lambda dp\ p^2 \bigg(\frac{\tanh \left(\frac{1}{2} \beta  \omega_-(p,\Delta) \right)}{\omega_-(p,\Delta)}
+\frac{\tanh \left(\frac{1}{2} \beta  \omega_+(p,\Delta) \right)}{\omega_+(p,\Delta)}\bigg)\ ,
\end{equation}
Solving this equation at $\Delta=0$ one can obtain the value of the critical temperature
at which the phase transition takes place.

\medskip
If one is interested in studying electromagnetic properties, such as the
Meissner effect, one must include in \eqref{bcsl} a $U(1)$ gauge field,
$A_\mu$. This gauge field is going to be treated as an external field.
 
One can also consider fluctuations of the gap, 
$\Delta(\vec x)=\Delta_0+\bar\Delta(\vec x)$, 
around its equilibrium position, $\Delta_0$, determined by eq.~\eqref{gapBCS}.
This will allow us to compute the coherence length, which is a measure of the
rigidity of the condensate. For simplicity, we will consider static
fluctuations and suppose $\Delta$ to be real.
 
We are going to treat the inclusion of the gauge field and the fluctuations of
the gap as static perturbations. To this purpose, we  write the Lagrangian
\eqref{bcsl} as $\mathcal L=\Psi^\dagger O_F\Psi$, where 
$\Psi^\dagger=(\psi^\dagger,\psi_c^\dagger)$, and we split the matrix $O_F$ as
$O_F(\Delta,A)=O_{F0}(\Delta_0)+\delta O_F(\bar\Delta,A)$.
In the path integral formalism this amounts to consider the saddle point
approximation, which is the approach that Gor'kov \cite{Gorkov} followed to 
derive the Ginzburg-Landau effective action from the BCS theory. 
In this way, the free energy can be expanded as
\begin{equation}\label{expansion}
\Omega=\int d^3x \,V_{\rm cl}(\Delta_0+\bar\Delta(\vec x))\ -\frac{1}{2\beta}\log\det O_F=\Omega_0+\Omega_1+\Omega_2+\ldots
\end{equation}
\begin{align*}
\Omega_0&=\int d^3x\, g^2\Delta_0^2\ -\frac{1}{2\beta}\log\det O_{F0}\ ,\\
\Omega_1&=\int d^3x\, 2g^2\Delta_0\bar\Delta(\vec x)\ -\frac{1}{2\beta}{\rm Tr}[O_{F0}^{-1}\delta O_F]\ ,\\
\Omega_2&=\int d^3x\, g^2\bar\Delta^2(\vec x)\ +\frac{1}{4\beta}{\rm Tr}[(O_{F0}^{-1}\delta O_F)^2]\ ,
\end{align*}
where 
\begin{align}
{\rm Tr}[O_0^{-1}\delta O]&\equiv\int d^4x\int d^4x_1\,{\rm tr}[O_0^{-1}(x,x_1)\delta O(x_1,x)]\ ,\\
{\rm Tr}[(O_0^{-1}\delta O)^2]&\equiv\int d^4x\int d^4x_1\int d^4x_2\int d^4x_3\,{\rm tr}[O_0^{-1}(x,x_1)
\delta O(x_1,x_2)O_0^{-1}(x_2,x_3)\delta O(x_3,x)]\ ,
\end{align}
$\delta O(x,y)=\delta^{(4)}(x-y)\delta O(x)$ and $\rm tr[.]$ is the usual
trace over matrix elements. The first term of the expansion \eqref{expansion},
$\Omega_0$, is just the free energy corresponding to eq.~\eqref{Veff},
which fixes the value of the gap.
The second term, $\Omega_1$, has two contributions, one corresponding to 
$\bar\Delta$  (which vanishes, as it is proportional to the gap equation) 
and another one due to the gauge field $A$. 
In momentum space we have
\begin{equation}
{\rm Tr}[O_{F0}^{-1}\delta O_F]=\int d^4x\, \frac{1}{\beta}\sum_n\int\frac{d^3p}{(2\pi)^3}{\rm tr}
[O_{F0}^{-1}(\omega_n,\vec p)\delta O_F(x)]\ ,
\end{equation}
which gives rise to terms of the form
\begin{equation}
\int d^4x\, \frac{1}{\beta}\sum_n\int\frac{d^3p}{(2\pi)^3}h(\omega_n,p^2)\vec p\cdot \vec A(x)\ ,
\end{equation}
for some function $h$ depending on $\omega_n$ and $p^2$. This term does not survive the momentum
integration, so this just leaves contributions involving the temporal component of the gauge field,
which are interpreted as fluctuations or space inhomogeneities of the chemical
potential.
 
Once the effective potential is computed we have to add the kinetic term for
the gauge field, the complete free energy is then
\begin{align}\label{GL}
\Omega_{\rm tot}&=\int d^3x\,\left[ \frac{1}{2}\vec E^2
+\frac{1}{2}\vec B^2\right]+\Omega(\Delta,A)
\nonumber\\&=
\int d^3x\,\left[f_1\Big\vert\left(\vec\nabla-ie\vec A\right)\Delta\Big\vert^2
+m^{-2}\vert\bar\Delta\vert^2+f_2\vec E^2+f_3\vec B^2\right]
+\Omega_0+\ldots\ ,
\end{align}
where the coefficients $f_1$, $f_2$, $f_3$ and $m^{-2}$ have been introduced
to account for the contributions coming from the $\Omega_1$ and $\Omega_2$
terms of the expansion of the free energy. Near $T_c$, we can expand
$\Omega_0$ in eq.~\eqref{GL} as
\begin{equation}\label{ab}
\Omega_0(\Delta_0,T)=\int d^3x\,
\Big[a(T-T_c)\Delta_0^2+b\Delta_0^4+\ldots\Big]\ ,
\end{equation}
where
\begin{align*}
a&=\frac{g^4}{2\pi^2}\beta_c^2\int_{0}^\Lambda dp\ p^2\left({\rm sech}^2\left(\frac{1}{2} \beta_c(\omega_0(p)+\mu )\right)
+(\mu\rightarrow-\mu)\right)\ ,\\
b&=\frac{g^8}{2\pi^2}\int_{0}^\Lambda dp\ p^2\left(-\beta_c\frac{{\rm sech}^2
\left(\frac{1}{2} \beta_c(\omega_0(p)+\mu )\right)}{(\omega_0(p)+\mu )^{2}} 
+2\frac{\tanh\left(\frac{1}{2} \beta_c  (\omega_0(p)+\mu )\right)}{(\omega_0(p)+\mu )^3}
+(\mu\rightarrow-\mu)\right)\ .
\end{align*}
Eq. \eqref{ab} is just the Ginzburg-Landau (GL) free energy, which is valid near the critical 
temperature, so \eqref{GL} must be considered as a generalization of the GL
free energy, valid for all temperatures well below the cut-off scale.

From the equations of motion for $\vec A$ (in the London limit, i.e. 
$\bar \Delta=0$, and in the Coulomb gauge, $\vec\nabla\cdot\vec A=0$) one sees
that the magnetic field is exponentially suppressed inside the 
superconductor, which is the footprint of the Meissner effect. The
corresponding magnetic penetration length, $\lambda$, is given by
\begin{equation}\label{lambdam}
\nabla^2\vec A=\frac{e^2f_1}{f_3}\vert\Delta_0\vert^2\vec A\qquad
\Rightarrow\qquad \frac{1}{\lambda^2}=\frac{e^2f_1}{f_3}\vert\Delta_0\vert^2\ . 
\end{equation}

Another characteristic length is the coherence length, $\xi$, which
characterizes the distance over which the superconducting electron 
concentration cannot drastically change. 
This length is obtained from the equations of motion for $\bar\Delta(x)$ 
obtained from \eqref{GL}. In absence of gauge field, we get
\begin{equation}\label{coherencelength}
\nabla^2\bar\Delta=\frac{1}{m^2f_1}\bar\Delta\qquad\Rightarrow
\qquad\frac{1}{\xi^2}=\frac{1}{m^2f_1}\ .
\end{equation}

Hence to determine these lengths we must be able to identify the 
coefficients $f_1$, $f_3$ and $m^{-2}$ in \eqref{GL}. The identification of
these  coefficients is explained in appendix~\ref{ap:coefficients}. 

We also want to compute the specific heat and the critical 
magnetic field above which superconductivity is destroyed. The former was 
already computed in \cite{Barranco-Russo} at constant $\mu$, 
instead of constant charge density. This way is more convenient because,
when neutral scalars are considered in the SUSY case, they do not contribute to
the charge density constraint $\rho=d\Omega/d\mu$. The specific heat is computed through
the formula
\begin{equation}\label{c}
S=-\left(\frac{\partial V_{\rm eff}}{\partial T}\right)_\mu\,,\quad c=T\left(\frac{dS}{dT}\right)_\mu
=-T\left(\frac{\partial^2V_{\rm eff}}{\partial T^2}+\frac{\partial^2V_{\rm eff}}{\partial T\partial\varepsilon}
\frac{\partial\varepsilon}{\partial T}\right)\,, \quad\text{where}\quad\frac{\partial\varepsilon}{\partial T}
=-\frac{\partial_T\partial_\varepsilon V_{\rm eff}}{\partial_\varepsilon^2V_{\rm eff}}\ .
\end{equation}
The critical magnetic field is obtained by equating the work done for holding the magnetic 
field out of the superconductor with the condensation energy, 
\begin{equation}\label{Hc}
\frac{H_c^2(T)}{8\pi}=V_n(T)-V_s(T)\ ,
\end{equation}
where $V_n$ and $V_s$ are the free energies per unit volume in the normal and 
superconducting phase at zero field. This is the critical magnetic field above 
which superconductivity would be destroyed in a type I superconductor, where 
the GL parameter is $\kappa\equiv\nicefrac{\lambda}{\xi}\ll 1$. 
When $\kappa\gg 1$, we are dealing with type II superconductors.
Instead of a  discontinuous breakdown of superconductivity in a first order
phase transition at $H_c$, like in type I superconductors, type II
superconductors exhibit an intermediate Abrikosov vortex state in between two
critical magnetic fields,
\begin{equation}
H_{c1}\approx \frac{\phi_0}{2\pi\lambda^2}\ ,\qquad\qquad H_{c2}\approx\frac{\phi_0}{2\pi \xi^2}\ .
\end{equation}
$H_{c1}$ is the value of the magnetic field for 
which a single vortex with flux quantum $\phi_0=\nicefrac{\pi}{e}$ would 
appear, whereas near $H_{c2}$ vortices are as closely packaged as the 
coherence length allows.
In this vortex state the magnetic
field penetrates in regular arrays of flux tubes of non-superconducting
material surrounded by a superconducting current. Superconductivity
disappears completely for magnetic fields above $H_{c2}$.

\section{Supersymmetric BCS theory}

The supersymmetric version of BCS theory \cite{Barranco-Russo}, 
is based on an $\mathcal N=1$ SUSY theory described by the following K\"ahler 
potential, 
\begin{equation}\label{Kahler}
K=\Phi^\dagger\Phi +g^2(\Phi^\dagger \Phi)^2\ ,
\end{equation}
and no superpotential. In \cite{Barranco-Russo} it was necessary to include up to four 
chiral superfields. Two superfields with the mentioned K\"ahler potential for 
each one, in order to have Dirac fermions, just as in standard BCS theory; and 
two additional free superfields with canonical K\"ahler potential to ensure that the 
$U(1)_R$ symmetry is non-anomalous, so that we can introduce a chemical potential
for this symmetry in a consistent way.

The chemical potential is introduced in this theory for the $U(1)_R$ symmetry 
in such a way that scalars belonging to chiral superfields with K\"ahler potential
\eqref{Kahler} are neutral and, as a consequence, their fermionic superpartners have 
$R$-charge $-1$. This is done in this way in order to prevent these scalars from coupling 
to the chemical potential, thus avoiding Bose-Einstein condensation (BEC) of the scalars. 
However, the additional superfields, added to ensure that the $U(1)_R$ symmetry is 
non-anomalous, must contain fermions with $R$-charge $+1$ and therefore scalars 
with $R$-charge $+2$, which couple to the chemical potential and condense. 
Nevertheless, the four chiral superfields are decoupled form each other, 
then the sector with scalars suffering from BEC 
does not participate in the thermodynamics of the sector with K\"ahler 
potential  \eqref{Kahler}, in which we are interested.
For this reason, it is sufficient to study the one superfield model \eqref{Kahler}.

In this minimal setup, with 
one chiral superfield, this particular choice of $R$-charges prevents the 
introduction  of a superpotential, which may lead to mass terms, in this way 
we are dealing with a massless theory.  

This theory also has a baryonic $U(1)_B$ symmetry. If the chemical potential 
is introduced for this $U(1)_B$ symmetry instead for the $U(1)_R$ symmetry, 
this would inevitably lead to BEC for the scalar in the presence of a Fermi 
surface. In this case, there is no reason why we cannot introduce a 
superpotential that generates mass terms, for example $W=m\Phi\Phi$. With this 
superpotential we obtain the following scalar energy eigenvalues,
\begin{equation}
\omega_{S\pm}=\sqrt{p^2+m^2+4g^4\Delta^2}\pm\mu\ .
\end{equation}
As we have seen, the existence of a Fermi surface requires $\mu>m$, for which 
$\omega_{S-}$ would become negative at low momentum near the 
non-superconducting phase ($\Delta=0$), making the thermal contribution to the 
effective potential ill-defined, resulting in the appearance of BEC.

Once the K\"ahler potential \eqref{Kahler} is expanded in terms of component 
fields, we have to perform, as in the previous section, the 
Hubbard-Stratonovich transformation introducing the auxiliary field $\Delta$,
which now is related with the fermion condensate through
\begin{equation}
\Delta=\frac{\psi\psi}{1+4g^2v^2}\ .
\end{equation}
After expanding the scalar field to quadratic order around its vacuum 
expectation value, $v$, which turns out to be zero, going to Euclidean space 
and introducing the chemical potential for the $R$-symmetry we arrive at the 
following Lagrangian
\begin{equation}\label{SUSYBCS}
\mathcal L_{E}=\left(\partial^a\phi\partial^a\phi^*
+\bar\psi \bar \sigma^0\left(\partial^\tau+\mu\right)\psi
+i\bar\psi \bar \sigma^i\partial^i\psi\right)
+4g^4\vert\Delta\vert^2\vert\phi\vert^2-g^2\Delta(\psi\psi)-g^2\Delta^*(\bar\psi\bar\psi)\ ,
\end{equation}
with classical potential $V_{\rm cl}=g^2\vert\Delta\vert^2$.

Now we want to study the magnetic response of the superconductor, therefore we 
have to  turn on an external  $U(1)$ gauge field. We have two possibilities:
the gauge field can be turned on for the baryonic $U(1)_B$ symmetry or the 
$U(1)_R$ symmetry%
\footnote{$R$-symmetry can only be gauged within the context of supergravity. 
The present model can be easily embedded in $\mathcal N=1$ supergravity, in 
such a way that in the usual laboratory set up, where energy configurations 
are much lower than the Planck scale,  the supergravity multiplet can be 
ignored.}. In either case supersymmetry is broken by the background.  
In principle, the magnetic response could depend on this choice, but as 
explained at the end of appendix~\ref{ap:coefficients}, this is not the case, 
at least for a small enough gauge coupling%
\footnote{However, things are different if we turn on the gauge fields
associated with the $U(1)_B$ and $U(1)_R$ symmetries at the same time. In this 
case, there is a linear combination of the previously introduced gauge fields 
which defines a massless rotated gauge field. Therefore, the magnetic field
associated to this rotated gauge field can penetrate the superconductor 
without being subject to the Meissner effect. This case is similar to that of 
 \cite{Ferrer-Incera,Manuel}, so it would be interesting to study how the 
phenomenology presented there translates into the 
present supersymmetric model.}. For definiteness we introduce the 
gauge field through the baryonic $U(1)_B$ symmetry, therefore the Lagrangian 
\eqref{SUSYBCS} is modified to 
\begin{align}\label{SUSYgaugeL}
\mathcal L_{E}&=\left(\partial^a\phi\partial^a\phi^*
+\bar\psi \bar \sigma^0\left(\partial^\tau-ieA^\tau+\mu\right)\psi
+i\bar\psi \bar \sigma^i\left(\partial^i-ieA^i\right)\psi\right)
\nonumber\\&\phantom{=\ }
-ieA^a\phi\partial^a\phi^*+ieA^a\phi^*\partial^a\phi
+e^2A^aA^a\vert\phi\vert^2+4g^4\vert\Delta\vert^2\vert\phi\vert^2
-g^2\Delta(\psi\psi)-g^2\Delta^*(\bar\psi\bar\psi)\ .
\end{align}
Given this Lagrangian, we can construct the bosonic and fermionic matrices 
(whose explicit form  is written in appendix \ref{ap:matrices}) and split them 
as explained in the previous section, so that we have to add to 
eq.~\eqref{expansion}, the scalar contribution
\begin{equation}
+\frac{1}{2\beta}\log\det O_{S0}+\frac{1}{2\beta}
{\rm Tr}[O_{S0}^{-1}\delta O_S]-\frac{1}{4\beta}
{\rm Tr}[(O_{S0}^{-1}\delta O_S)^2]+\ldots
\end{equation}
The determination of the quantities explained in the case of 
relativistic BCS theory is now completely analogous for the supersymmetric 
case. For example, the effective potential analogous to \eqref{Veff}, becomes 
now
\begin{align}\label{VeffSUSY}
V_{\rm eff}&=g^2\Delta_0^2
\nonumber\\&\phantom{=\ }
-\frac{1}{2}\int\frac{d^3p}{(2\pi)^3}(\omega_++\omega_--2\omega_S)
\nonumber\\&\phantom{=\ }
-\frac{1}{\beta}\int\frac{d^3p}{(2\pi)^3}\left(\log(1+e^{-\beta\omega_+})
+\log(1+e^{-\beta\omega_-})-2\log(1-e^{-\beta\omega_S})\right)\ ,
\end{align}
where the energy eigenvalues are
\begin{equation}
\omega_{F\pm}=\sqrt{(p\pm \mu)^2+4g^4\Delta_0^2}\ ,\qquad\qquad\omega_{S\ 1,2}=\sqrt{p^2+4g^4\Delta_0^2}\ .
\end{equation}

\section{Comparison between the SUSY model and  relativistic BCS}

\subsection{Gap}
 
According to the previous section, we have to solve the gap equation 
$\partial_{\varepsilon}V_{\rm eff}=0$. Just as shown in \cite{Barranco-Russo}, 
the explicit form of the gap equation is 
\begin{align}\label{gappa}
1 &=\frac{g^2 }{2\pi ^2}\int_0^\Lambda dp\ p^2 \bigg(\frac{\tanh \left(\frac{1}{2} \beta  \sqrt{4g^4 \Delta_0^2+(p-\mu )^2}\right)}
{\sqrt{4g^4 \Delta_0^2 +(p-\mu )^2}}+\frac{\tanh \left(\frac{1}{2} 
\beta \sqrt{4g^4\Delta_0^2 +(p+\mu )^2}\right)}
{\sqrt{4 g^4 \Delta_0^2 +(p+\mu )^2}}
\nonumber\\&\phantom{=\ }
-\frac{2 \coth \left(\frac{1}{2} \beta \sqrt{4g^4\Delta_0^2 +p^2}\right)}{\sqrt{4g^4 \Delta_0^2 +p^2}}\bigg)\ ,
\end{align}
where the second line is the new contribution due to the scalar.
\begin{figure}[t]
\centering
\begin{tabular}{cc}
\setlength{\unitlength}{1mm}
\begin{picture}(82,50)
%
%
\put(0,0){\includegraphics[width=0.46\textwidth]{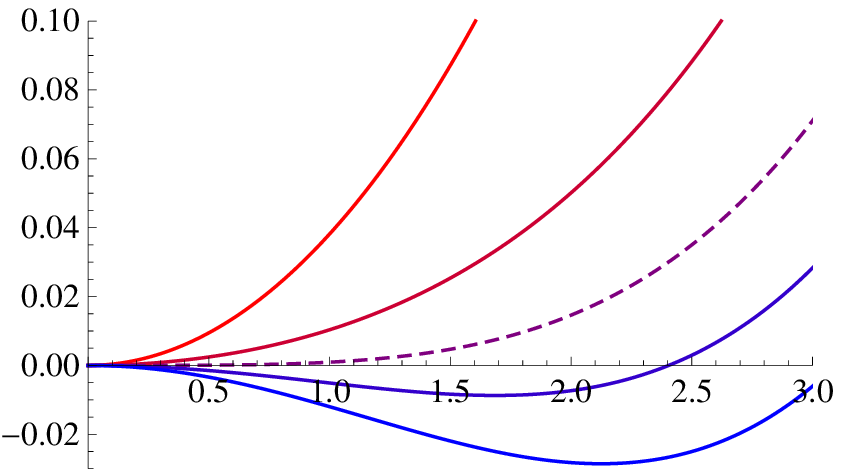}}
\put(75.5,34){\footnotesize $T_{c}$}
\put(78.5,9){\small $\Delta_0$}
\put(5,46){\small $V_{eff}$}
\end{picture}
&
\setlength{\unitlength}{1mm}
\begin{picture}(82,50)
%
%
\put(0,0){\includegraphics[width=0.43\textwidth]{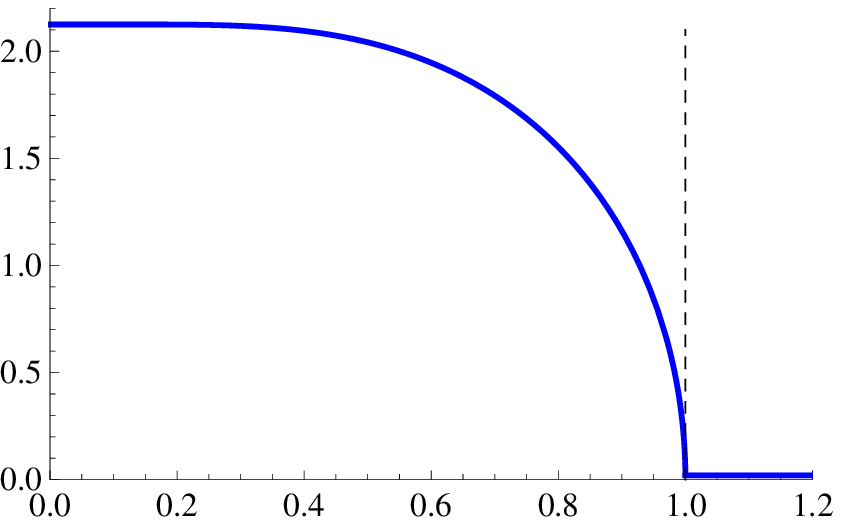}}
\put(58.5,45){\footnotesize $T_{c}$}
\put(75,3){\small $T$}
\put(2,46){\small $\Delta_0$}
\end{picture}
\\
(a)&(b)\\[1mm]
\setlength{\unitlength}{1mm}
\begin{picture}(82,50)
%
%
\put(0,0){\includegraphics[width=0.45\textwidth]{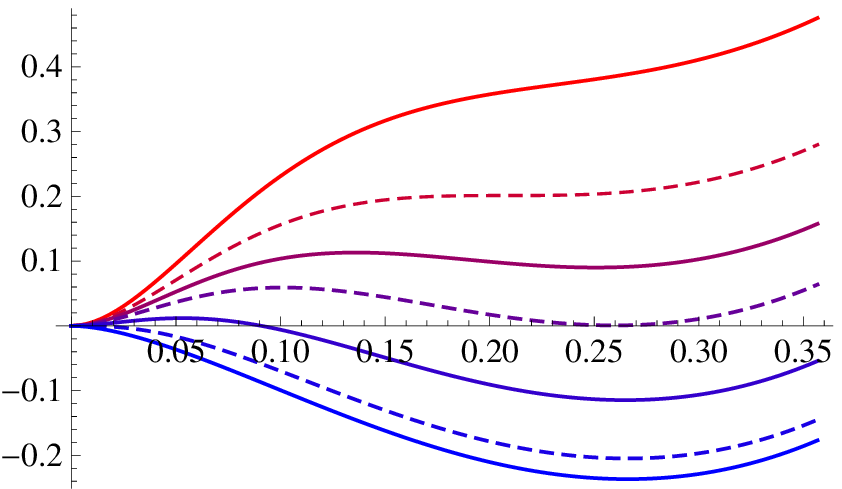}}
\put(80.5,6.5){\makebox[0pt][r]{\footnotesize $T_{c1}$}}
\put(80.5,19.2){\makebox[0pt][r]{\footnotesize $T_{c2}$}}
\put(80.5,32){\makebox[0pt][r]{\footnotesize $T_{c3}$}}
\put(78.5,14.1){\small $\Delta_0$}
\put(3,47){\small $V_{eff}$}
\end{picture}&
\setlength{\unitlength}{1mm}
\begin{picture}(82,50)
%
%
\put(0,0){\includegraphics[width=0.42\textwidth]{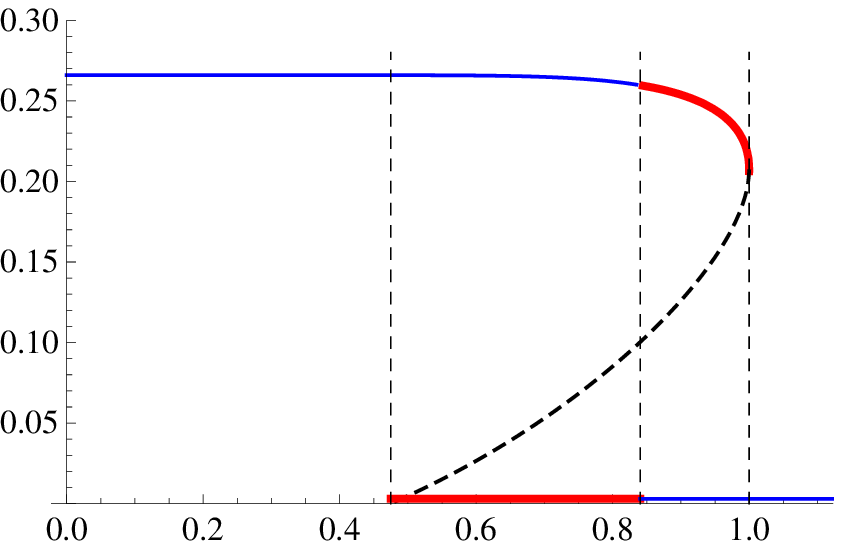}}
\put(31.2,45){\footnotesize $T_{c1}$}
\put(52.5,45){\footnotesize $T_{c2}$}
\put(62,45){\footnotesize $T_{c3}$}
\put(14.5,42){\footnotesize zone 1}
\put(43,42){\footnotesize 2}
\put(59,42){\footnotesize 3} 
\put(68,42){\footnotesize 4}
\put(74,3){\small $T$}
\put(4,47.5){\small $\Delta_0$}
\end{picture}\\
(c)&(d)
\end{tabular}
\caption{(a) Effective potential as function of the gap for different 
temperatures ($T=2,\,1.3,\,1,\,0.75,\,0$ from top to bottom) in the relativistic BCS case. (b) Corresponding gap as function 
of the temperature ($\mu=0.0065$, $g=0.54$, $\Lambda=8.4$).
(c) ($T=1.1,\,1,\,0.92,\,0.84,\,0.7,\,0.47,\,0$) and (d) analogous figures for the SUSY case 
($\mu=0.65$, $g=3.9$, $\Lambda=52$). The thicker red lines in (d) correspond to metastable 
solutions and the curved dashed line represents the maximum that separates 
both minimums in the effective potential. From now on, all figures and data are given in units of $T_c$.}\label{gap}
\end{figure} 
For the standard relativistic BCS case the effective potential develops a 
non-trivial minimum at $\Delta_0\neq 0$, which will approach the origin as we 
increase the temperature reaching the zero gap value at the critical 
temperature  $T_c$, (fig.~\ref{gap} (a) and (b)). The phase transition
is second order.
For the SUSY case, at low temperatures, $T<T_{c1}$, the effective potential 
has a unique non-trivial minimum (zone 1 of fig.~\ref{gap} (c) and (d)), above 
$T_{c1}$ (zone 2) a zero gap metastable minimum appears becoming the dominant 
one for $T>T_{c2}$  (zone 3), so that the minimum with non-vanishing gap 
becomes metastable until it disappears at $T_{c3}$, jumping to the 
non-superconducting zero gap solution for $T>T_{c3}$ (zone 4). Therefore,
the phase transition is first order in the supersymmetric case.
For short, we will be calling $T_c\equiv T_{c3}$ and 
$\Delta_c\equiv\Delta(T_{c3})$ from now on.

As explained in \cite{Barranco-Russo}, the IR physics of massless scalars makes 
the phase transition first order rather than second order. First order phase 
transitions in usual superconductivity have already 
been explained in \cite{Halperin}, where a gauge field takes a non zero vacuum 
expectation value appearing as a $\langle A^2\rangle\vert\Delta_0\vert^3$ term 
in the free energy, which inevitably leads to a first order phase transition. 
This gauge field plays the same role as the scalar in the supersymmetric case. 
Near $\Delta_0=0$ and at low momentum, we can approximate the contribution of 
the scalar as  
\begin{equation}
\partial_\varepsilon V_{\rm eff}\approx\frac{2g^4T}{\pi^2}
\int_0dp\,p^2\frac{1}{(p^2+4g^4\Delta_0^2)}
\approx-\frac{2g^6T}{\pi}\Delta_0\ ,
\end{equation}
which leads to the analogous $\mathcal O(\vert\Delta_0\vert^3)$ term in the 
free energy.

\medskip
At low temperatures, we can neglect in \eqref{gappa} the antiparticle and the 
scalar contributions for momenta near the Fermi surface, where the main 
contribution to the integral comes from. By doing so, we are taking the 
non-relativistic limit and connecting with the standard BCS result.
Specifically, one performs the \textit{ad hoc} approximation by which one 
substitutes $dp^3$ by $4\pi p_F^2dp$ and the integral is done in the interval 
$\vert p-p_F\vert <\Lambda$ around the Fermi surface. After these 
approximations and at large cut-off ($\Lambda/T_c\gg 1$), one can 
show that the behaviour of the gap near the critical temperature $T_c$  follows 
a universal behaviour
\begin{equation}\label{D(T)}
\frac{\Delta_{0}(T)}{\Delta_0(0)}\approx\eta\sqrt{1-\frac{T}{T_c}}\ ,
\end{equation}
where $\eta=1.74$ can be computed numerically from eq.~\eqref{ab}. In the 
supersymmetric case we obtain  numerically the following behaviour near the 
critical temperature:
\begin{equation}\label{D(T)SUSY}
\frac{\Delta_0(T)-\Delta_c}{\Delta_0(0)}
\approx \eta \left(1-\frac{T}{T_c}\right)^\alpha\ ,
\end{equation}
where $\alpha\approx 0.5$ and now
$\eta$ depends on the parameters $g$, $\mu$ and $\Lambda$.

\medskip
We can also determine the expressions for the gap at zero temperature. 
The non-relativistic and relativistic BCS expressions are:
\begin{equation}\label{gapLambda}
\vert\Delta_0(0)_{\rm BCS}\vert\approx \frac{\Lambda}{g^2}
e^{-\frac{2\pi^2}{g^2\mu^2}}
\approx\frac{1}{2}\pi e^{-\gamma}\frac{T_c}{g^2}\ ,
\qquad\qquad
\vert\Delta_0(0)_{\rm rel.\,BCS}\vert\approx \frac{\Lambda^3}{6\pi^2}
\approx \frac{T_c}{g^2}\ ,
\end{equation}
where in the non-relativistic limit we have considered the aforementioned approximation
and $\gamma$ is the Euler-Mascheroni constant. Both the gap at zero temperature and the 
critical temperature depend on the cut-off (the Debye energy in the standard BCS case), 
in such a way that the cut-off dependence disappears from the formula for the gap once 
it is expressed in terms of the critical temperature. We have also shown here the 
dependence on the cut-off because we want to compare 
with the supersymmetric case, since one of the virtues of supersymmetry is
the softening of divergences due to cancellations between fermionic and bosonic
contributions. As the thermal integrals in \eqref{Veff} and 
\eqref{VeffSUSY} are convergent, it is sufficient to study the 
dependence of the gap with the cut-off at zero temperature.
Comparing the relativistic  expression with the supersymmetric one,
\begin{equation}\label{gapLambdaSUSY}
\vert\Delta_0(0)_{\rm SUSY\,BCS}\vert\approx \frac{\Lambda}{g^2} e^{-\frac{\pi^2}{g^2\mu^2}-\frac{3}{2}}\ ,
\end{equation}
we observe that this softening in the cut-off dependence  appears in the present system.
However, this reduction in the power of the cut-off does not appear if we compare the supersymmetric 
expression with the non-relativistic one, where the dependence  is linear rather than cubic
as in the relativistic case. This is because the power of the cut-off has already been reduced
after performing the substitution $dp^3\rightarrow 4\pi p_F^2dp$ in the non-relativistic case. 
Finally, just as in the non-supersymmetric cases \eqref{gapLambda}, one can check that the supersymmetric 
case \eqref{gapLambdaSUSY} is also linear with the critical temperature with no dependence on the cut-off.

\subsection{Specific heat}

\begin{figure}[t]
\centering
\begin{tabular}{cc}
\setlength{\unitlength}{1mm}
\begin{picture}(82,50)
%
%
\put(0,0){\includegraphics[width=0.44\textwidth]{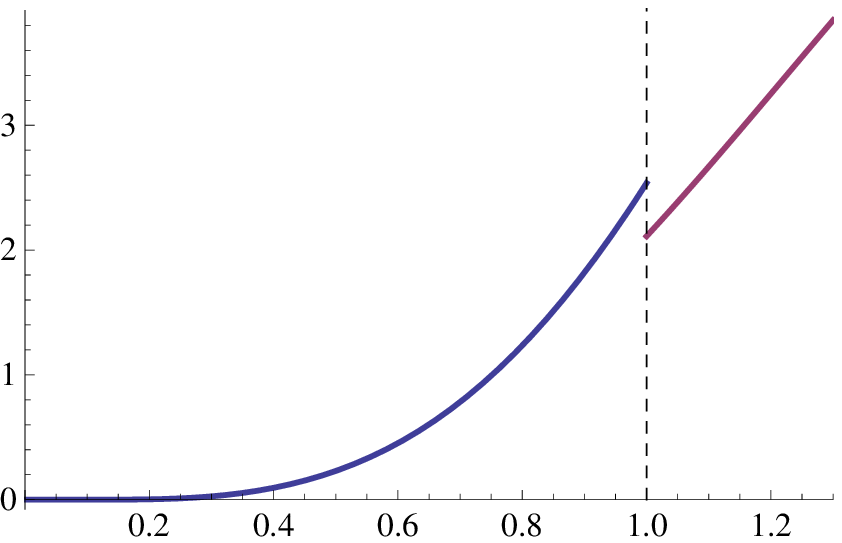}}
\put(77,3){\small $T$}
\put(1.5,49){\small $c$}
\end{picture}&
\setlength{\unitlength}{1mm}
\begin{picture}(82,50)
%
%
\put(0,0){\includegraphics[width=0.44\textwidth]{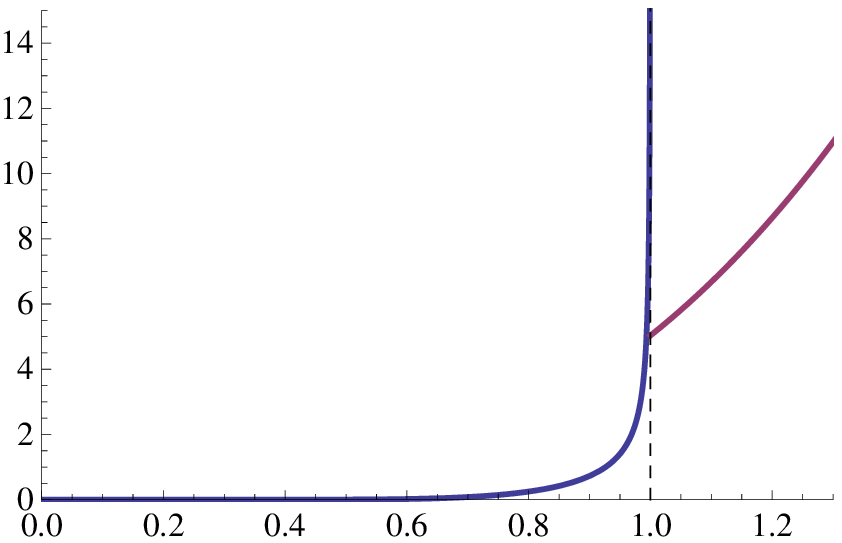}}
\put(77,3){\small $T$}
\put(3,49){\small $c$}
\end{picture}
\\
(a)&(b)
\end{tabular}
\caption{Specific heat a function of the temperature. 
(a) Relativistic BCS ($\mu=0.0065$, $g=0.54$, $\Lambda=8.4$). (b) SUSY BCS ($\mu=0.65$, $g=3.9$, $\Lambda=52$).}\label{specificheat}
\end{figure}

The comparison between the SUSY and the relativistic BCS case for the specific heat, computed  with eq.~\eqref{c}, is shown in fig.~\ref{specificheat}.
When only fermions are considered the jump in the specific heat at the 
critical temperature is finite, as it is characteristic for second order phase 
transitions, whereas it is infinite when the scalar is included. 

The behaviour of the specific heat in the different temperature regimes is 
explained in \cite{Barranco-Russo}, whose expressions we show 
here%
\footnote{There is a $\nicefrac{1}{2}$ factor between the results presented
here from those of \cite{Barranco-Russo}, because here we are considering just 
one complex scalar field and a Weyl fermion.}. At high temperatures 
($\Delta_0=0$) the different contributions to the specific heat corresponding 
to the scalar, the fermion and its anti-particle, are the following ones
\begin{equation}
c_S\vert_{\Delta_0=0}=\frac{4\pi^2T^3}{15}\ ,\qquad\qquad c_{F\pm}\vert_{\Delta_0=0}=\frac{7\pi^2T^3}{60}\ ,
\qquad \text{for\ }T\gg\mu\ .
\end{equation}

At low temperatures the antiparticle and the scalar can be neglected and 
the main contribution comes from the region near the Fermi surface $p\sim\mu$, 
so that the behaviour of the specific heat in the normal phase and in the 
superconducting phase is given by
\begin{equation}
c_{F-}\vert_{\Delta_0=0}\sim\frac{\mu^2T}{6}\ ,\qquad\qquad 
c_{F-}\vert_{\Delta_0\neq0}\sim e^{-\frac{2g^2\Delta_0(T=0)}{T}}\ .
\end{equation}
The last expression shows a way to compute the value of the gap at zero 
temperature. Indeed, by performing a fit of the plots in 
fig.~\ref{specificheat} one can obtain the value of the gap at zero 
temperature shown in fig.~\ref{gap}.

\subsection{Magnetic penetration length and coherence length}

\begin{figure}[h!]
\centering
\begin{tabular}{cc}
\setlength{\unitlength}{1mm}
\begin{picture}(82,50)
%
%
\put(0,0){\includegraphics[width=0.42\textwidth]{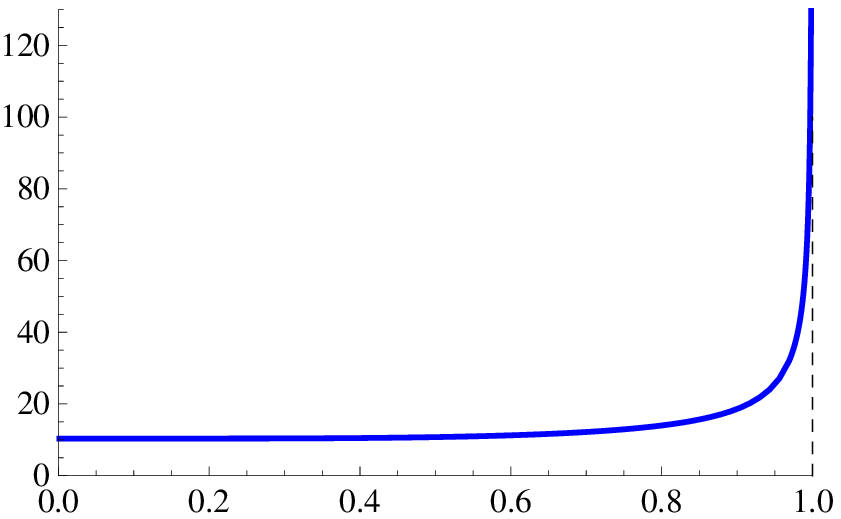}}
\put(74.5,3){\small $T$}
\put(3,47){\small $e\lambda$}
\end{picture}&
\setlength{\unitlength}{1mm}
\begin{picture}(82,50)
%
%
\put(0,0){\includegraphics[width=0.42\textwidth]{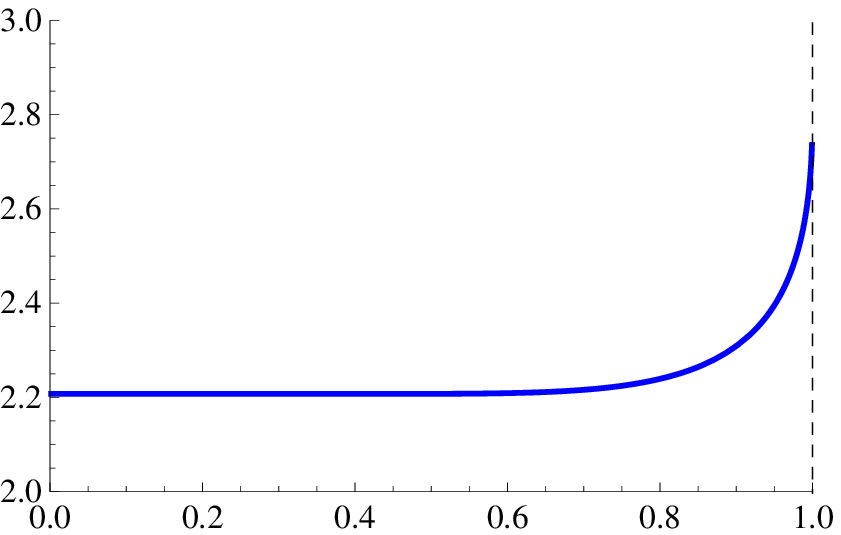}}
\put(74.5,3){\small $T$}
\put(2.2,47){\small $e\lambda$}
\end{picture}
\\
(a)&(b)
\end{tabular}
\caption{Magnetic penetration length as a function of the temperature.  (a)~Relativistic BCS ($\mu=0.0065$, $g=0.54$, $\Lambda=8.4$). 
(b)~SUSY BCS ($\mu=0.65$, $g=3.9$, $\Lambda=52$). }\label{mlength}
\end{figure}

The magnetic penetration length, obtained from eq.~\eqref{lambdam}, is plotted in fig.~\ref{mlength}.
In both the relativistic and the SUSY BCS theory, the magnetic penetration 
length is a monotonically increasing function%
\footnote{For high enough values of the chemical potential and the cut-off a 
counterintuitive non-trivial minimum can appear before reaching the critical 
temperature, which would mean that there is a range of temperatures where the 
Meissner effect is enhanced with increasing temperature. This odd behaviour 
can be avoided by restricting the parameter  range of validity 
to not very high values of the chemical potential and the cut-off. However, 
this restriction is relaxed in the SUSY case where the cut-off dependence is 
softened.}.

For the fermionic contribution alone, the magnetic penetration length diverges 
near the critical temperature. Whereas in the supersymmetric case, it reaches 
a finite value at $T_c$  and it jumps to infinity for $T>T_c$, which is the 
expected behaviour for a first order phase transition. This behaviour is 
explained basically from the gap dependence. Expanding the $f_1$ coefficient
as a power series of the temperature and the gap, one obtains
\begin{equation}\label{f1expansion}
f_{1\rm\ rel. BCS}=a_0+a_1\Delta_0^2 +a_2(T-T_c)+\ldots\ ,\qquad\qquad 
f_{1\rm\ SUSY\,BCS}=\alpha_0+\alpha_1(\Delta_0^2-\Delta_c^2)+\ldots\ ,
\end{equation}
where one can check that the coefficients shown do not vanish. The ellipsis 
stands for higher powers of the temperature, taking into account the 
temperature dependence of the gap, \eqref{D(T)} or \eqref{D(T)SUSY}.
Substituting this expansion in \eqref{lambdam}, and using \eqref{D(T)} or \eqref{D(T)SUSY}, 
one finds the  behaviour
\begin{equation}
\lambda_{\rm rel.\, BCS}\sim \left(1-\frac{T}{T_c}\right)^{-1/2}\ ,
\qquad\qquad 
\left(\lambda-\lambda_c\right)_{\rm SUSY\, BCS}\sim
\left(1-\frac{T}{T_c}\right)^\alpha\ . 
\end{equation}
 
The behaviour of the magnetic penetration length at zero temperature can be 
computed analytically. According to the dependence of the coefficients $f_1$ 
and $m^{-2}$ with the cut-off, given in appendix~\ref{ap:coefficients}, 
and that of the gap, we find the following expressions for the 
magnetic penetration length, 
\begin{equation}
\lambda_{\rm rel.\,BCS}\approx\frac{4g^3}{3\sqrt{3}\pi^2e}\Lambda^2\ ,\qquad 
\lambda_{\rm SUSY\,BCS}\approx\frac{2\pi}{ec} (1+4c^2)^{3/4}\Lambda^{-1}\ ,\qquad 
\text{where}\  c=\exp\left[{-\frac{\pi^2}{g^2\mu^2}-\frac{3}{2}}\right]\ . 
\end{equation}

\begin{figure}[h!]
\centering
\begin{tabular}{cc}
\setlength{\unitlength}{1mm}
\begin{picture}(82,50)
%
%
\put(0,0){\includegraphics[width=0.42\textwidth]{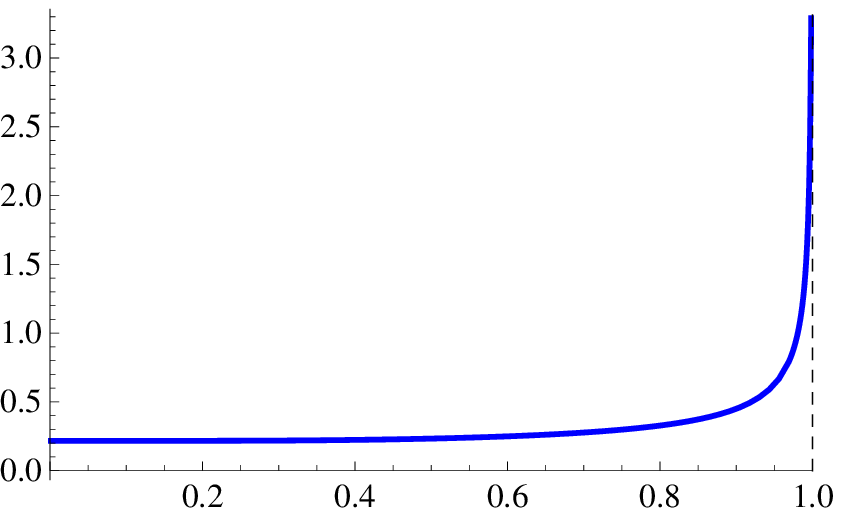}}
\put(74.5,3){\small $T$}
\put(3.5,47){\small $\xi$}
\end{picture}
&
\setlength{\unitlength}{1mm}
\begin{picture}(82,50)
%
%
\put(0,0){\includegraphics[width=0.42\textwidth]{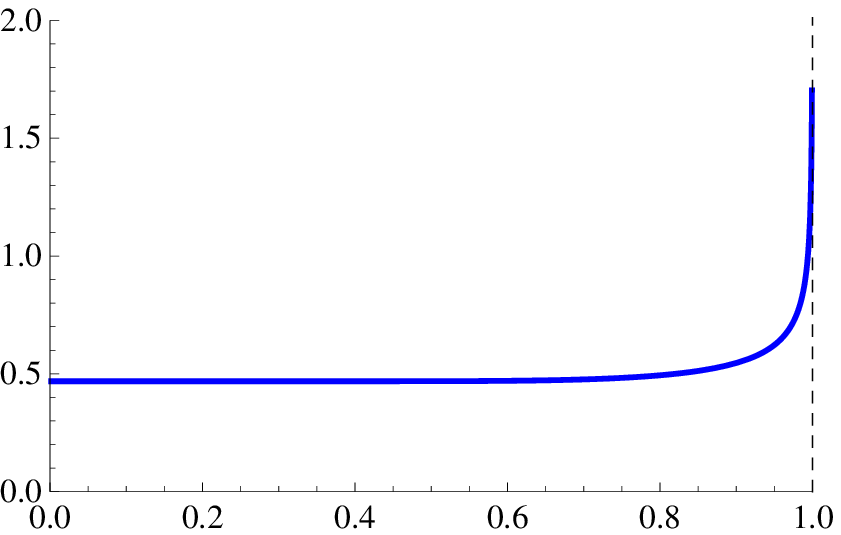}}
\put(74.5,3){\small $T$}
\put(4,47){\small $\xi$}
\end{picture}
\\
(a)&(b)
\end{tabular}
\caption{Coherence length as a function of the temperature. 
(a)~Relativistic BCS ($\mu=0.0065$, $g=0.54$, $\Lambda=8.4$). (b)~SUSY BCS ($\mu=0.65$, $g=3.9$, $\Lambda=52$).}\label{clength}
\end{figure}

\medskip
The coherence length, $\xi$, obtained from eq.~\eqref{coherencelength}, is 
plotted in fig.~\ref{clength}. It is a monotonically increasing function of 
the temperature for both the relativistic and the SUSY BCS theory.
The behaviour near the critical temperature is the same as for the magnetic 
penetration length. To see this, expand the $m^{-2}$ coefficient as we did 
with the $f_1$ coefficient. Using the gap equation, one can see that the 
$m^{-2}$ coefficient has a global $\Delta_0^2$ factor so that the expansions 
are  
\begin{equation}\label{m2expansion}
m^{-2}_{\rm rel.\, BCS}=\Delta_0^2(b_0 + b_1\Delta_0^2+b_2(T-T_c)+\ldots)\ ,
\qquad\qquad
m^{-2}_{\rm SUSY\, BCS}=\Delta_0^2(\beta_0 
+\beta_1(\Delta_0^2-\Delta_c^2)+\ldots)\ .
\end{equation}
Inserting the expansions \eqref{f1expansion} and \eqref{m2expansion} in the 
expression for the coherence length, \eqref{coherencelength},
we find 
\begin{equation}
\xi_{\rm rel.\, BCS}\sim \left(1-\frac{T}{T_c}\right)^{-1/2}\ ,
\qquad\qquad 
\left(\xi-\xi_c\right)_{\rm SUSY\, BCS}
\sim\left(1-\frac{T}{T_c}\right)^\alpha\ .
\end{equation}
Thus, the coherence length exhibits the same behaviour as the magnetic 
penetration length. However, the behaviour of the coherence length at zero 
temperature is different from that of the magnetic penetration length. Now,
the dependence on the cut-off is
\begin{gather}
\xi_{\rm rel.\,BCS}\approx\frac{9\sqrt{6}\pi^4}{4g^4}\Lambda^{-5}\ ,\\ 
\xi_{\rm SUSY\,BCS}\approx\frac{g}{2(1+4c^2)^{3/4}}
\left(2\pi^2+g^2\mu^2 \left(\frac{32c^4+16c^2+5}{(1+4c^2)^{5/2}}
-2\log\frac{1+\sqrt{1+4c^2}}{2c}\right)\right)^{-1/2}\!\!+\mathcal O(\Lambda^{-2})\ .
\end{gather}

\begin{figure}[h!]
\centering
\begin{tabular}{cc}
\setlength{\unitlength}{1mm}
\begin{picture}(82,50)
%
%
\put(0,0){\includegraphics[width=0.42\textwidth]{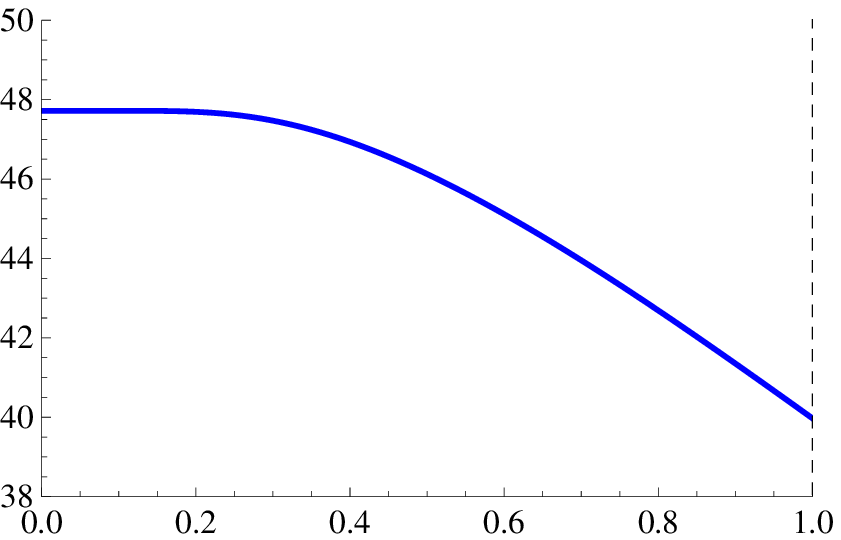}}
\put(74.5,3){\small $T$}
\put(2.6,48){\small $e\kappa$}
\end{picture}
&
\setlength{\unitlength}{1mm}
\begin{picture}(82,50)
%
%
\put(0,0){\includegraphics[width=0.42\textwidth]{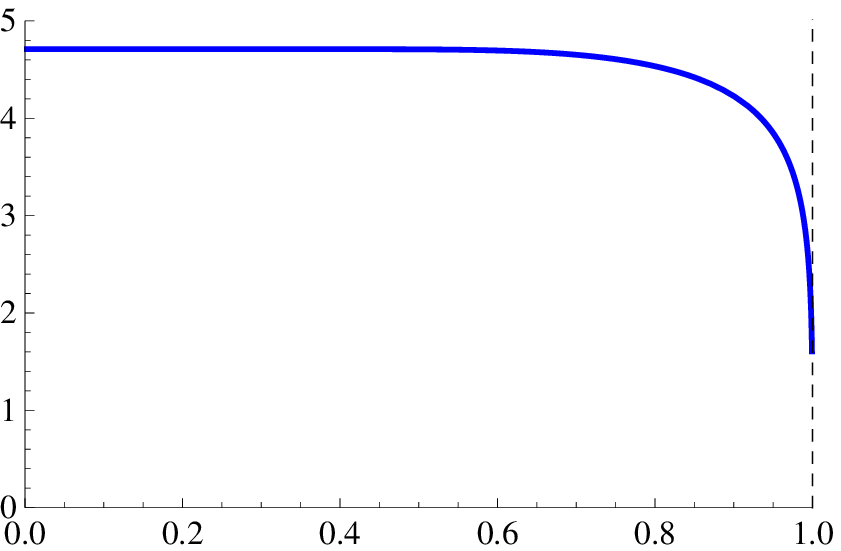}}
\put(74.5,3){\small $T$}
\put(1,48){\small $e\kappa$}
\end{picture}
\\
(a)&(b)
\end{tabular}
\caption{Ginzburg-Landau parameter as a function of temperature. 
(a)~Relativistic BCS ($\mu=0.0065$, $g=0.54$, $\Lambda=8.4$). (b)~SUSY BCS ($\mu=0.65$, $g=3.9$, $\Lambda=52$).}\label{kappa}
\end{figure}

\medskip
If we take the quotient between these two characteristic lengths,
\begin{equation}
\kappa=\frac{\lambda}{\xi}=\frac{1}{\sqrt{2}e\,m\,f_1\Delta_0}\ ,
\end{equation}
we get the Ginzburg-Landau parameter, shown in fig.~\ref{kappa}. As the 
coherence length behaves in the same way as the magnetic penetration length 
near the phase transition, at leading order, the GL parameter will take a 
finite constant value. Depending on the value of the GL parameter one has a 
type I  ($\kappa\ll1$) or a type II superconductor ($\kappa\gg 1$). In the GL 
theory $\kappa$ is defined near the critical temperature and the critical 
value differentiating between the two types of superconductor is 
$\kappa=1/\sqrt{2}$. As shown in fig.~\ref{kappa}, $\kappa\gg1$,
since we are considering  small values of gauge coupling $e$, then the 
superconductors are type II in both the relativistic BCS case and the SUSY 
case.

\subsection{Critical magnetic fields}

For a type I superconductor, the critical magnetic field is obtained by equating the energy per unit 
volume, associated with holding the field out against the magnetic pressure, 
with the condensation energy. That is eq.~\eqref{Hc},
\begin{equation*}
\frac{H_c^2(T)}{8\pi}=V_n(T)-V_s(T)\ .
\end{equation*}

\begin{figure}[h!]
\centering
\begin{tabular}{cc}
\setlength{\unitlength}{1mm}
\begin{picture}(78,54)
%
%
\put(5,5){\includegraphics[width=0.42\textwidth]{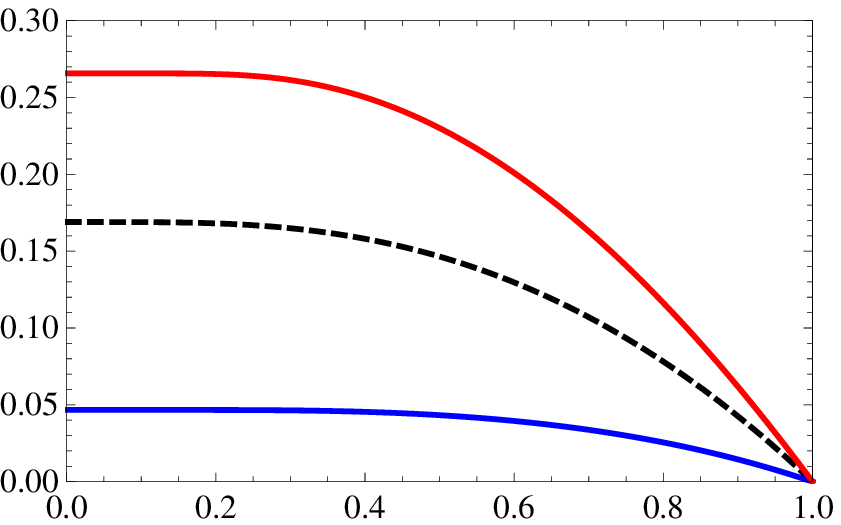}}
\put(-2,43){\small\makebox[0pt][l]{$H_{c2}$}}
\put(-2,30){\small\makebox[0pt][l]{$H_{c}$}}
\put(-2,14.5){\small\makebox[0pt][l]{$H_{c1}$}}
\put(41,1){\small $T$}
\end{picture}
&
\setlength{\unitlength}{1mm}
\begin{picture}(78,54)
%
%
\put(5,5){\includegraphics[width=0.42\textwidth]{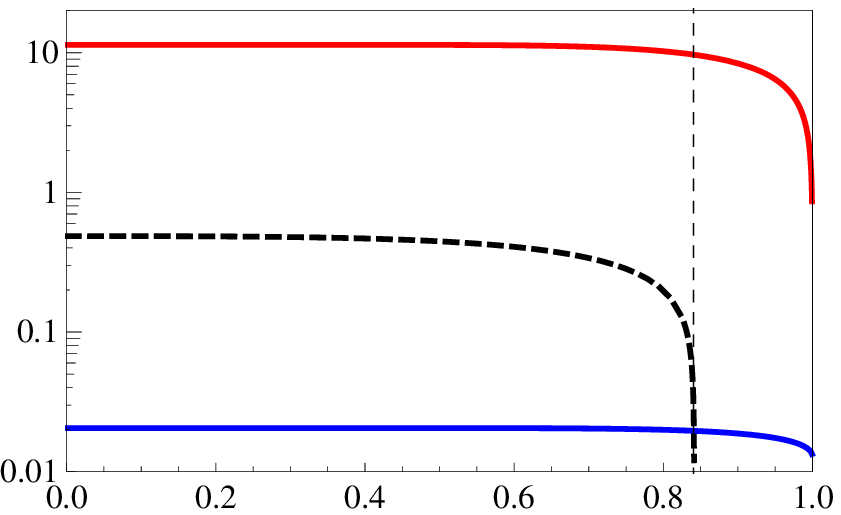}}
\put(0,44){\small\makebox[0pt][l]{$H_{c2}$}}
\put(0,28){\small\makebox[0pt][l]{$H_{c}$}}
\put(0,12){\small\makebox[0pt][l]{$H_{c1}$}}
\put(43,1){\small $T$}
\end{picture}\\
(a)&(b)
\end{tabular}
\caption{Critical magnetic fields $H_c$ (black), $H_{c1}$ (blue) and $H_{c2}$ 
(red) as a function of the temperature.  
(a) Relativistic BCS ($e=0.2$, $\mu=0.0065$, $g=0.54$, $\Lambda=8.4$), for clarity $H_{c1}$ and $H_{c2}$ have been 
rescaled by a factor $50$ and $0.005$ respectively). 
(b) SUSY BCS ($e=0.2$, $\mu=0.65$, $g=3.9$, $\Lambda=52$).} 
\label{Hs}
\end{figure}

The behaviour near the critical temperature is found by preforming expansions 
similar to those for the $f_1$ and $m^{-2}$ coefficients. In the standard 
relativistic BCS theory the gap is expanded around $\Delta_0=0$. On the other 
hand, when the scalar is considered, $H_c$ does not make sense above $T_{c2}$,
since the superconducting minimum in the effective potential becomes 
metastable, but we can perform the expansion around  
$\Delta_{c2}\equiv\Delta_0(T_{c2})$. According to the dependence of the gap 
with the temperature \eqref{D(T)} in the relativistic BCS theory and due to 
the fact that $\Delta_0$ is linear with the temperature near $T_{c2}$ in the 
SUSY BCS theory, we have
\begin{align}
\frac{H_{c\ \rm rel.\,BCS}^2}{8\pi}&=
\partial_T\partial_\varepsilon(V_n(T)-V_s(T))
\Big\vert_{\substack{T=T_c\\\Delta_0=0}}(T-T_c)\Delta_0^2
\nonumber\\&\phantom{=\ }
+\frac{1}{2}\partial^2_\varepsilon(V_n(T)-V_s(T))
\Big\vert_{\substack{T=T_c\\\Delta_0=0}}\Delta_0^4+
\ldots\\
\frac{H_{c\ \rm SUSY\,BCS}^2}{8\pi}&=
\partial_T(V_n(T)-V_s(T))
\Big\vert_{\substack{\,\,T=T_{c2}\\\Delta_0=\Delta_{c2}}}(T-T_{c2})+\ldots 
\end{align}
from which we find a linear and square root behaviour near the critical 
temperatures, 
\begin{equation}
H_{c\ \rm rel.\,BCS}\sim \left(1-\frac{T}{T_c}\right)\ ,
\qquad\qquad 
H_{c\ \rm SUSY\,BCS}\sim\sqrt{1-\frac{T}{T_{c2}}}\ ,
\end{equation}
as  shown in fig.~\ref{Hs} (a) and (b), respectively. The expressions for the  
critical magnetic field at zero temperature are
\begin{equation}
H_{c\ \rm rel.\,BCS}\approx\frac{\sqrt{2\pi}g}{3\pi^2}\Lambda^3\ ,
\qquad 
H_{c\ \rm SUSY\,BCS}\approx\sqrt{8\pi} \sqrt{-1-\frac{ g^2\mu^2}{\pi^2} 
\left(\frac{1}{\sqrt{1+4c^2}}
- \log\frac{1+\sqrt{1+4c^2}}{2c}\right)}\frac{c}{g}\Lambda\ .
\end{equation}
As expected the dependence on the cut-off is milder in the 
supersymmetric case.  

As explained at the end of sec.~\ref{sec:2}, for type II superconductors, 
which are characterized by the appearance of Abrikosov vortices
in a mixed superconductor-normal state, there are two critical magnetic 
fields, $H_{c1}$ and $H_{c2}$, 
\begin{equation*}
H_{c1}\approx\frac{\phi_0}{2\pi\lambda^2}\ ,\qquad\qquad H_{c2}\approx\frac{\phi_0}{2\pi\xi^2}\ .
\end{equation*} 
They are plotted in fig.~\ref{Hs} together with $H_c$.
It is easy to obtain the behaviour of these two critical magnetic fields in 
the different regimes  using the expressions for the magnetic penetration 
length and the coherence length. 

If $H_c\ll H_{c2}$, we will be able to see the
intermediate vortex state as we decrease the applied magnetic field, i.e. the 
superconductor is type II. On the contrary if $H_c\gg H_{c2}$, we reach the 
pure superconducting state without the formation of any vortex, and we have 
type I superconductivity. According to fig.~\ref{Hs}, we have type II 
superconductivity in both the relativistic and the SUSY
BCS theory, in agreement with the prediction obtained in the previous section 
by computing the GL parameter.

Given that $H_c$ ends at $T_{c2}$  instead of $T_{c3}$, as $H_{c1}$ and 
$H_{c2}$, and that there is a range of parameters where $H_{c2}<H_c$ at zero 
temperature, one can find a crossing between the two magnetic fields and a 
crossover between type I and type II behaviour as we increase the temperature. 
However, this crossing effect disappears if the gauge coupling is sufficiently small. 

\section{Conclusions}

In this paper  we have computed some relevant quantities in superconductivity 
as a function of the temperature for the supersymmetric BCS model proposed in 
\cite{Barranco-Russo} and for standard relativistic BCS theory, comparing both 
results with the aim of clarify the role of supersymmetry. 
These quantities  are the gap, the specific heat, the magnetic penetration 
length, which describes the Meissner effect, the coherence length and the 
different critical magnetic fields.  To compute these 
quantities we have considered spatial fluctuations of the gap and introduced 
an external gauge field for the baryonic $U(1)_B$ symmetry.  Both the gap 
fluctuations and the gauge field are introduced as perturbations, so once the 
effective potential is expanded and the coefficients in the generalized 
Ginzburg-Landau equation \eqref{GL} are identified, the computation of the 
aforementioned quantities is straightforward.

Considering the quotient between the magnetic penetration length and the 
coherence length, which defines the GL parameter, shows that the studied 
superconductors are type II. This is confirmed when studying the relation 
between the critical magnetic fields $H_c$ and $H_{c2}$.  

The main differences between the standard relativistic BCS results and those 
from the supersymmetric model lie in the different orders of the phase 
transition. For example, the infinite jumps found in the supersymmetric model 
for the specific heat, magnetic penetration length and coherence length are 
those characteristic of a first order phase transition. Another important 
difference is the softening in the cut-off dependence in all quantities thanks 
to supersymmetry. 

It would be very interesting to see if
there are realistic condensed matter systems
with an effective quasi supersymmetric dynamics with thermodynamic
and response properties similar
to the present system, summarized in the different figures.
It would also be interesting to construct extensions of the present
supersymmetric model, in particular,
to describe supersymmetric color superconductivity and compare the
resulting dynamics with the  standard phenomenology of QCD color
superconductivity \cite{Iwasaki-Iwado,Alford-Rajagopal-Wilczek,Rapp-Schafer-Shuryak-Velkovsky}.

\section*{Acknowledgements}

I am very grateful to J. G. Russo for his supervision during this project and 
his valuable advise and suggestions. I also acknowledge financial support by 
MECD FPU Grant No AP2009-3511 and projects FPA 2010-20807 and Consolider CPAN.


\appendix

\section{Scalar and Fermionic matrices}\label{ap:matrices}

Writing the quadratic part of the Lagrangian in the presence of a baryonic 
gauge field, \eqref{SUSYgaugeL}, as 
\begin{equation*}
\mathcal L=\Phi^\dagger O_S\Phi+\Psi^\dagger O_F\Psi\ ,\qquad\text{with}\qquad 
\Phi=\left(
\begin{array}{c}
\phi(p)\\
\phi^*(-p)
\end{array}
\right)\qquad\text{and}\qquad
\Psi=\left(
\begin{array}{c}
\psi_1(p)\\
\psi_2(p)\\
\psi^\dagger_1(-p)\\
\psi^\dagger_2(-p)
\end{array}
\right)\ ,
\end{equation*}
the scalar and fermionic matrices after the splitting 
$O(\Delta,A)=O_0(\Delta_0)+\delta O(\bar\Delta,A)$,
are given in momentum space by
\begin{gather}
O_{S0}=\left(
\begin{array}{cc}
 \frac{1}{2}\left(\omega ^2+p^2\right)+2g^4\Delta_0^2 & 0 \\
 0 & \frac{1}{2}\left(\omega ^2+p^2\right)+2g^4\Delta_0^2 \\
\end{array}
\right)\ ,\nonumber\\
\qquad\delta O_S=\left(
\begin{array}{cc}
 \parbox{5cm}{\centering $e(A_\tau\omega +\vec A\cdot\vec p)+\frac{1}{2}
 e^2\left(A_\tau^2+\vec A^2\right)$\\$+2g^4 \left(2\bar \Delta\Delta_0 +\bar \Delta^2\right)$} & 0 \\
 0 &  \parbox{6cm}{\centering $-e(A_\tau\omega +\vec A\cdot\vec p)+\frac{1}{2}
 e^2
 \left(A_\tau^2+\vec A^2\right)$\\$+2g^4 \left(2\bar \Delta\Delta_0 +\bar \Delta^2\right)$}
\end{array}
\right)\ ,\nonumber\\
O_{F0}=\left(
\begin{array}{cccc}
 \frac{i}{2}\omega +\frac{1}{2}p+\frac{1}{2}\mu & 0 & 0 & -g^2\Delta_0 \\
 0 & \frac{i}{2}\omega -\frac{1}{2}p+\frac{1}{2}\mu & g^2 \Delta_0 & 0 \\
 0 & g^2\Delta_0 & \frac{i}{2}\omega +\frac{1}{2}p-\frac{1}{2}\mu & 0 \\
 -g^2 \Delta_0 & 0 & 0 & \frac{i}{2}\omega -\frac{1}{2}p-\frac{1}{2}\mu \\
\end{array}
\right)\ ,\nonumber\\
\delta O_F=\left(
\begin{array}{cccc}
- \frac{i}{2}eA_\tau-\frac{1}{2}eA & 0 & 0 & -g^2\bar  \Delta  \\
 0 &-\frac{i}{2}eA_\tau+\frac{1}{2}eA & g^2\bar  \Delta  & 0 \\
 0 & g^2\bar  \Delta  & \frac{i}{2}eA_\tau+\frac{1}{2}eA & 0 \\
 -g^2\bar \Delta  & 0 & 0 &  \frac{i}{2}eA_\tau-\frac{1}{2}eA \\
\end{array}
\right)\ .
\label{matrixB}
\end{gather}

If instead of introducing a baryonic $U(1)_B$ gauge field, we had introduced
$U(1)_R$ gauge field, the scalar and fermionic matrices would have been the 
following ones:
\begin{gather}
\delta O_S=\left(
\begin{array}{cc}
 2g^4 \left(2\bar \Delta\Delta_0 +\bar \Delta^2\right) & 0 \\
 0 & 2g^4 \left(2\bar \Delta\Delta_0 +\bar \Delta^2\right)
\end{array}
\right)\ ,\nonumber\\
\delta O_F=\left(
\begin{array}{cccc}
\frac{i}{2}eA_\tau+\frac{1}{2}eA & 0 & 0 & -g^2\bar  \Delta  \\
 0 &\frac{i}{2}eA_\tau-\frac{1}{2}eA & g^2\bar  \Delta  & 0 \\
 0 & g^2\bar  \Delta  & -\frac{i}{2}eA_\tau-\frac{1}{2}eA & 0 \\
 -g^2\bar \Delta  & 0 & 0 &  -\frac{i}{2}eA_\tau+\frac{1}{2}eA \\
\end{array}
\right)\ .
\label{matrixR}
\end{gather}
The energy eigenvalues computed for the $O_{S0}$ and $O_{F0}$ matrices are
\begin{equation}
\omega_{F\pm}=\sqrt{(p\pm \mu)^2+4g^4\Delta_0^2}\ ,\qquad\qquad\omega_{S\ 1,2}=\sqrt{p^2+4g^4\Delta_0^2}\ .
\end{equation}

\section{$m^{-2}$, $f_1$  and $f_3$ coefficients}\label{ap:coefficients}

In momentum space, the $m^{-2}$ and $f_1$ terms in \eqref{GL} are given by
\begin{align}
\int d^3x\,m^{-2}\bar\Delta\bar\Delta^*&=\int \frac{d^3k}{(2\pi)^3}\,m^{-2}
\bar\Delta^*(\vec k)\bar\Delta(\vec k)\ ,\\
\int d^3x\,f_1\partial^i\Delta\,\partial^i\Delta^*&=\int \frac{d^3k}{(2\pi)^3}\,f_1
\vec k^2\bar\Delta^*(\vec k)\bar\Delta(\vec k)\ ,
\end{align}
where we have considered time independent perturbations.
Thus, we have to find in \eqref{expansion} a term quadratic in 
$\bar\Delta$ and expand its coefficient up to quadratic order in momentum. 
The zero order term will correspond to $m^{-2}$ and the coefficient of the
quadratic term in momentum will be identified with $f_1$. 
Terms quadratic in $\bar\Delta$ are 
found in $\Omega_2$ and, if scalars are considered, in $\Omega_1$. 
The $\Omega_1$ term only contributes to the $m^{-2}$ coefficient, which in 
momentum space becomes 
\begin{equation}
\frac{1}{2\beta}{\rm Tr}\left[O_{S0}^{-1}\delta O_S\right]=
\frac{1}{2\beta^2}\int d^4x\sum_n\int\frac{d^3K}{(2\pi)^3}{\rm tr}\left[O_{S0}^{-1}(\omega_n,\vec K)\delta O_S(\vec x)\right]\ ,
\end{equation}
where we have to sum over Matsubara frequencies,  $\omega_n=2n\pi /\beta$ for 
bosonic frequencies and $\omega_n=(2n+1)\pi/\beta$ for fermionic ones. Once 
the Matsubara sums are done, we have to consider the piece quadratic in 
$\bar\Delta$ (supposing $\bar\Delta$ to be real)
\begin{align}
\frac{1}{2\beta}{\rm Tr}\left[O_{S0}^{-1}\delta O_S\right]_{\bar\Delta\bar\Delta}&=
\frac{1}{\beta^2}\int d^4x\int\frac{d^3K}{(2\pi)^3}\mathfrak B_{\bar\Delta\bar\Delta}(\vec K)\bar\Delta(\vec x)\bar\Delta(\vec x)\nonumber\\
&=\frac{1}{\beta}\int\frac{d^3K}{(2\pi)^3}\mathfrak B_{\bar\Delta\bar\Delta}(\vec K)\int \frac{d^3k}{(2\pi)^3}\bar\Delta^{*}(\vec k)\bar\Delta(\vec k)\ .
\end{align}
Hence, we identify the first contribution to $m^{-2}$ as
\begin{equation}
m^{-2}=\frac{1}{\beta}\int\frac{d^3K}{(2\pi)^3}
\mathfrak B_{\bar\Delta\bar\Delta}(\vec K)+\ldots
\end{equation}
Let us elaborate now on the $\Omega_2$ contribution,
\begin{equation}
\frac{1}{4\beta}{\rm Tr}[(O_{F0}^{-1}\delta O_F)^2]=\frac{1}{4\beta}\int d^4x_1\int d^4x_2\,{\rm tr}
[\delta O_F(\vec x_1)O_{F0}^{-1}(x_1,x_2)\delta O_F(\vec x_2)O_{F0}^{-1}(x_2,x_1)]\ ,
\end{equation}
plus  the analogous scalar term if one considers the supersymmetric case, and 
the extra term $\int d^3x\,g^2\bar\Delta^2$ for $m^{-2}$. The previous expression in momentum space is
\begin{align}\label{mathcal F}
\frac{1}{4\beta}{\rm Tr}[(O_{F0}^{-1}\delta O_F)^2]&=\frac{1}{4\beta^3}\int d^4x_1d^4x_2\sum_{m,n}
\int \frac{d^3k_1}{(2\pi)^3}\frac{d^3k_2}{(2\pi)^3}\frac{d^3q_1}{(2\pi)^3}\frac{d^3q_2}{(2\pi)^3}
\nonumber\\&\hspace*{4cm}
\times e^{-i(\omega_m-\omega_n)(\tau_1-\tau_2)}
e^{-i(\vec k_1-\vec k_2+\vec q_1)\vec x_1}e^{-i(-\vec k_1+\vec k_2+\vec q_2)\vec x_2}
\nonumber\\&\hspace*{4cm}
\times{\rm tr}[\delta O_F(\vec q_1)O_{F0}^{-1}(\omega_m,\vec k_1)
\delta O_F(\vec q_2)O_{F0}^{-1}(\omega_n,\vec k_2)]
\nonumber\\&=
\frac{1}{4\beta}\sum_{n}
\int\frac{d^3k_1}{(2\pi)^3}\frac{d^3k_2}{(2\pi)^3}
{\rm tr}[\delta O_F(\vec k_2-\vec k_1)O_{F0}^{-1}(\omega_n,\vec k_1)
\delta O_F(\vec k_1-\vec k_2)O_{F0}^{-1}(\omega_n,\vec k_2)]
\nonumber\\&\equiv\frac{1}{\beta}
\int\frac{d^3k_1}{(2\pi)^3}\frac{d^3k_2}{(2\pi)^3}
\mathcal F(\vec k_1,\vec k_2)\ .
\end{align}
Taking the piece quadratic in $\bar\Delta$ in  \eqref{mathcal F}, we have
\begin{align}
\Omega_{2}\big\vert_{\bar\Delta\bar\Delta}&=\frac{1}{\beta}\int \frac{d^3k_{1}}{(2\pi)^3}\frac{d^3k_{2}}{(2\pi)^3}\,
\bar\Delta^*(\vec k_2-\vec k_1)\bar\Delta(\vec k_2-\vec k_1)\mathcal F_{\bar\Delta\bar\Delta}(\vec k_1,\vec k_2)\ .
\end{align}
Assuming that second order corrections are located close to each other in 
momentum space, we can expand the momenta around their average value, 
$\vec k_1=\vec K-\nicefrac{\vec k}{2}$, $\vec k_2=\vec K+\nicefrac{\vec k}{2}$, so that the corresponding free energy term is
\begin{equation}
\Omega_{2}\big\vert_{\bar\Delta\bar\Delta}=\frac{1}{\beta}\int \frac{d^3K}{(2\pi)^3}\frac{d^3k}{(2\pi)^3}\,
\bar\Delta^*(\vec k)\bar\Delta(\vec k)\mathcal F_{\bar\Delta\bar\Delta}(\vec  K,\vec k)\ ,
\end{equation}
Expanding $\mathcal F_{\bar\Delta\bar\Delta}$ up to quadratic order in $k$, we 
identify the $m^{-2}$ and $f_{1}$ coefficients with
\begin{align}
m^{-2}&=g^2+\frac{1}{\beta}\int\frac{d^3K}{(2\pi)^3}
\big(\mathcal F_{\bar\Delta\bar\Delta}(\vec K,0)-\mathcal B_{\bar\Delta\bar\Delta}(\vec K,0)
+\mathfrak B_{\bar\Delta\bar\Delta}(\vec K)\big)\ ,\\
f_{1}&=\frac{1}{2\beta}\int\frac{d^3K}{(2\pi)^3}
\big(\partial_{k}^2\mathcal F_{\bar\Delta\bar\Delta}(\vec K,0)
-\partial_{k}^2 \mathcal B_{\bar\Delta\bar\Delta}(\vec K,0)\big)\ ,
\end{align}
once the analogous bosonic contribution is included.

\bigskip
The $f_3$-term in the Ginzburg-Landau free energy \eqref{GL} 
will have a contribution coming from the gauge field kinetic term 
plus contributions coming from the part of $\Omega_2$ quadratic in the gauge 
field, which will be proportional to the square of the gauge coupling, $e^2$, 
\begin{equation}
f_3=\frac{1}{2}+\mathcal O(e^2)\ .
\end{equation} 
As the gauge coupling is assumed to be small we can simply take 
$f_3=\nicefrac{1}{2}$. From this identification for the $f_3$ coefficient we 
see that there is no significant difference between turning on  a baryonic 
gauge field \eqref{matrixB} or an $R$-symmetry \eqref{matrixR}  gauge field,
since differences would appear to order $\mathcal O(e^2)$.

\bigskip
Once  these coefficients are computed, one can study their cut-off dependence 
in both the relativistic and supersymmetric case. We will restrict ourselves 
to the  zero temperature regime, where integrals can be performed 
analytically. In the zero temperature limit, the explicit forms of the 
coefficients $f_1$ and $m^{-2}$ are
\begin{align}\label{f1T0}
f_1\big\vert_{T=0}&=\frac{128 g^{12} \Delta_0^6 \mu 
+80 g^8 \Delta_0^4 \mu (\Lambda +\mu)^2+\mu^2 (\Lambda +\mu)^5
+2 g^4 \Delta_0^2 (\Lambda +\mu)^3 \left(3 \Lambda ^2+6 \Lambda\mu+8\mu^2\right)}
{96 \pi^2\Delta_0\left(4 g^4 \Delta_0^2+(\Lambda +\mu)^2\right)^{5/2}}
\nonumber\\&\phantom{=\ }
+(\mu \rightarrow -\mu )
\nonumber\\&\phantom{=\ }
+\frac{g^8 \Delta_0^2 \Lambda^3}{2 \pi^2 \left(4 g^4 \Delta_0^2+\Lambda^2\right)^{5/2}}\ ,
\end{align}

\begin{align}\label{m2T0}
m^{-2}\big\vert_{T=0}&=g^2+\frac{g^4}{4\pi^2}\left(
\frac{(\Lambda +\mu)(5\mu^2+2\Lambda\mu-\Lambda^2)+4 g^4 (5\mu-3\Lambda)\Delta_0^2}{\sqrt{4 g^4\Delta_0^2+(\Lambda +\mu )^2}}
\right.
\nonumber\\&\phantom{=\ }
+\left.2\left(\mu ^2-6 g^4 \Delta _0^2\right) 
\log\left(\frac{\mu +\sqrt{4 g^4 \Delta_0^2+\mu^2}}{\Lambda +\mu +\sqrt{4 g^4 \Delta_0^2+(\Lambda +\mu )^2}}\right) 
+(\mu \rightarrow -\mu)\right)
\nonumber\\&\phantom{=\ }
+\frac{g^4}{2 \pi^2}\left(\frac{12 g^4 \Delta_0^2 \Lambda +\Lambda^3}{\sqrt{4 g^4 \Delta_0^2+\Lambda^2}}
-12 g^4 \Delta_0^2{\rm csch}^{-1}\left(\frac{2 g^2\Delta_0}{\Lambda }\right)\right)\ ,
\end{align}
where the last line in \eqref{f1T0} or \eqref{m2T0} corresponds to the scalar 
contribution. Taking into account the cut-off dependence of the gap at zero 
temperature, \eqref{gapLambda},  we find the following expressions for the 
$f_1$ and $m^{-2}$ coefficients at leading order in $\Lambda$:
\begin{align}
f_{1\ \rm rel.\,BCS}&=\frac{3^5\pi^8}{8g^6}\Lambda^{-10}\ ,&
f_{1\ \rm SUSY\,BCS}&=\frac{g^4}{8\pi^2}(1+4c^2)^{-3/2}+\mathcal O(\Lambda^{-2})\ ,\\
m^{-2}_{\rm rel.\,BCS}&=g^2+\mathcal O(\Lambda^{-4})\ ,&
m^{-2}_{\rm SUSY\,BCS}&=g^2+\frac{g^4\mu^2}{2\pi^2} \left(\frac{32c^4+16c^2+5}{(1+4c^2)^{5/2}}
-2\log\frac{1+\sqrt{1+4c^2}}{2c}\right)\nonumber\\
&&&\hspace*{4mm}+\mathcal O(\Lambda^{-2})\ ,
\end{align}
where $c=\exp\left[-\frac{\pi^2}{g^2\mu^2}-\frac{3}{2}\right]$. We must stress 
that the coefficient $f_1$ in the relativistic BCS theory does not vanish at 
zero temperature, because $\Lambda$ is a physical cut-off acting like a ``Debye energy'', 
which  takes a finite value.


\end{document}